\documentstyle[12pt]{article}
\newcommand{\sect}[1]{\section{#1}\setcounter{equation}{0}}

\newcommand{\be}[1]{
\begin{equation}\label{#1}}
\newcommand{\ee}{\end{equation}}

\newcommand{\ba}{\begin{eqnarray}}

\newcommand{\nn}{\nonumber}
\newcommand{\ea}{\end{eqnarray}}

\newcommand{\res}[1]{{\rm Res}_{#1}}

\newcommand{\rra}{\ \longrightarrow \ }

\newcommand{\Map}[1]{\ \stackrel{#1}{\rra}}
\newcommand{\br }[1]{\left( #1 \right) }

\newcommand{\lbr }[1]{\left| #1 \right| }
\newcommand{\set }[1]{\left\{ #1 \right\} }
\newcommand{\re}[1]{~(\ref{#1})}

\newcommand{\Z}{{\bf Z}}

\newcommand{\R}{{\bf R}}

\newcommand{\bg}{{\cal H}^{\Gamma}}
\newcommand{\Bg}[1]{{\cal H}^{\Gamma_{ #1}}}
\newcommand{\bs}{{\cal H}^{\partial\Sigma}}
\newcommand{\Bs}[1]{{\cal H}^{\partial\Sigma_{ #1}}}
\newcommand{\cut}{\Sigma_{\Gamma}}

\newcommand{\bcut}{\Bs{\Gamma}}

\newcommand{\ev}{{\cal A}_{\Sigma}}
\newcommand{\Ev}[1]{{\cal A}_{\Sigma_{#1}}}
\newcommand{\dev}{{\cal A}^{\prime}_{\Sigma}}

\newcommand{\ecut}{\Ev{\Gamma}}
\begin{document}
\begin{titlepage}
\begin{flushright}
{\sc Report\ $\sharp$\ \ \ RU93-5-B}\\
\end{flushright}
\vspace{.6cm}

\begin{center}
{\Large{\bf DEFORMATION OF CFT BEYOND}}\\[5pt]
{\Large{\bf THE FIRST ORDER AND CLOSED}}\\[5pt]
{\Large{\bf STRING FIELD THEORY
\footnote{Talk given at the conference
{\em Journ\'{e}es Relativistes'93} held on April 5-7 1993 at
{\em l'Universit\'{e} Libre de Bruxelles}. To be published in the
{\em International Journal of Modern Physics D}.}}}\\[40pt]
{\sc Gregory Pelts}
\footnote{Supported by the DOE grant {\em DOE-91ER4651 Task B}}\\[2pt]
{\it
Department of Physics\\
The Rockefeller University\\
1230 York Avenue\\
New York, NY 10021-6399}\\[40pt]

{\sc Abstract}\\[12pt]

\parbox{13cm}{
A self-consistent string field theory with interaction is formulated. The
symmetry algebra of this theory includes, in the low-energy limit,
local space-time symmetries,  and the Brans-Dicke equation
describes a class  of low-energy solutions.}
\end{center}
\end{titlepage}

\sect{Introduction}
The situation in the string theory is inverse to that in
all other theories.
The Feynman rules, also called Polyakov rules,
are known, and physicists are trying to restore from them
a classical theory and symmetries.
This is important for decoupling of nonphysical
states, understanding the nonperturbative structure
and establishing connection with space-time geometry.
There is a belief that
classical closed string states can be associated with
quantum conformal field theories in two dimensions (CFT),
which are usually defined
as theories of the single string moving  in some nontrivial
space-time background. The condition of anomaly cancellation leads to the
so-called $\beta$-function equation on the background fields.
The main advantage of this approach is its more or less explicit connection
to space-time geometry and the main drawback that it usually focuses only on
massless fields.
Treatment of massive fields is problematic,
and, therefore, characterization of dynamical degrees of freedom is obscure.

Alternative approaches \cite{eo,sen,cnw,zw,rsz}
 are based on the operator  formalism \cite{alv}.
In \cite{eo,sen,cnw} a direct connection  has been found  between
(1,1)-primary fields of arbitrary mass level and deformations
of CFT.
Here, it will be shown how to generalize this result for treating of
finite deformations and to formulate the string field theory.

\sect{Axiomatic Conformal Field Theory}
The operator definition CFT \cite{alv} can be reformulated as
the following set of axioms:

\newtheorem{CFT}{Axiom}

\begin{CFT}
There is defined a functor ${\cal H}$ from the category of oriented
closed contours to the category of Hilbert spaces.
\end{CFT}
It means that to
any oriented contour $\Gamma$ there corresponds a Hilbert space $\bg$
and to any even map $\Gamma_1\Map{\nu}\Gamma_2$
there corresponds an isomorphism of the Hilbert spaces
$\Bg{1}\Map{\hat{\nu}}\Bg{2}$
in such a way that $\widehat{\nu_1\circ\nu_2}=\hat{\nu_1}\hat{\nu_2}$.
To a multicomponent contour $\Gamma=\bigcup_j \Gamma_j$ we will assign the
space $\bg =\bigotimes_j \Bg{j}$.
\begin{CFT}
The spaces corresponding to the contrary-oriented contours  are conjugated.
\end{CFT}
\begin{CFT}
To each bordered Riemann surface $\Sigma$ there corresponds in a conformal
invariant way a specific element $\ev$  of $\bs$ ({\em amplitude})
such that
\be{sewing}
\ev = Sp\br{\ecut}.
\ee
Here $\cut$ is a surface or set of surfaces resulting from the cutting of
$\Sigma$ along closed contour $\Gamma$,
and {\em Sp} is an operator contracting components of $\bcut$ corresponding
to two contrary oriented copies of $\Gamma$.
\end{CFT}
For the surface with the one and two component boundary,
$\ev$ can be considered to be the vacuum state and the propagator,
respectively.
To a multicomponent surface $\Sigma=\bigcup_j \Sigma_j$ we will assign the
amplitude  $\ev =\bigotimes_j \Ev{j}$.
\noindent
In order to describe CFT with central charge we should relax the axioms above,
changing them to their projective analogues.

\subsection{Vertex operators}
We will say that an element $\Psi$ of $\bs$  has
{\em support} in point $z_0\in\Sigma$,
if for any contour $\Gamma$ surrounding $z_0$ counter clockwise
it can be written as
$\Psi={\cal A}_{\Sigma_{\rm ext}}\Psi_0$, where $\Psi_0\in {\cal H}^{\Gamma}$
and $\Sigma_{\rm ext}$ is an external component of $\cut$.
Considering the map $\Psi\rra\Psi_0$ as an equivalence relation, we can
identify the spaces of states corresponding to different areas of the
Riemann surface but having the same support.
The Virasoro algebra can be shown to have a natural representation in the
space $H_{z_0}$ resulting from such identification.

The space $H_{z_1,\ldots,z_N}$ of states having  support in a set
of points $z_i$ is equivalent to $\bigotimes_{i=1}^N H_{z_i}$.
Elements of $H_{z_i}$ and the map
$H_{z_1}\times\ldots\times H_{z_N}\rra H_{z_1,\ldots,z_N}$
and  can be interpreted as vertex operators and their  $T$-product.

\sect{Deformation of CFT}
Let $\Psi$ be some vertex operator field (not necessarily primary).
If for deformed amplitude we will use the formula
\be{texp}
\dev =\mbox{Texp}\frac{1}{\pi}\int_{\Sigma}\Psi\, d^2\!z,
\ee
the condition\re{sewing} will be automatically satisfied
and only the condition of conformal invariance will remain to be implemented.

\subsection{Regularization}
We can not  use the simple {\em cutoff} regularization
$$
\int_{\Sigma}\phi d^2\! z\rra
\int_{\set{z\in\Sigma, \lbr{z-z_j}\geq r_j}}\phi d^2\! z
 $$
for integration of point-like contact singularities
in\re{texp} because it will violate the condition\re{sewing}.
Instead, we should use
an average of such regularization over cutoff parameters $r_j$
with the generalized measure $\mu$ in $\R_+$  defined by the formula
\be{reg}
\int_0^{\infty} r^{2\alpha} \mu (r)\,d\!r =\Lambda(\alpha)\ \ \br{\alpha\in\R}.
\ee
Here, $\Lambda$ is some smooth function on $\R$ satisfying $\Lambda(0)=1$
and $\Lambda(\alpha)=0$ for $\alpha$ bigger then some positive number.
A related proposal for regularization corresponding to specific
stepfunction $\Lambda=\Theta(-\alpha)d^{\alpha}$ was independently made in
\cite{rsz}.
It can be shown that the way in which the boundary singularity is
regularized
is not important as it does not affect the equivalence class of CFT.

\sect{Equation of Motion}
Let us identify deformed CFT vertex operators with initial CFT vertex
operators by means of the formula
$$
\Xi_{\Psi}=T\br{\Xi\exp{\br{\frac{1}{\pi}\int\Psi d^2 \!z}}}.
 $$
Deformed right and left components of energy momentum tensor can be shown
to be
\be{em}
{\cal T}^{\prime}=\br{{\cal T}+\Phi}_{\Psi}, \hspace{2em}
\bar{\cal T}^{\prime}=\br{\bar{\cal T}+\bar{\Phi}}_{\Psi},
\ee
where  $\Phi$, $\bar{\Phi}$ are some normalizable vertex operator functions.
The condition
\be{eq}
\bar{d} {\cal T}^{\prime}=d\bar{\cal T}^{\prime}=0
\ee
leads to dynamic equation on fields $\Psi$, $\Phi$, $\bar{\Phi}$
which can be interpreted as an equation of motion.
In fact, it is a conformal invariance condition for the deformed CFT.

\subsection{Symmetries}
It can be shown that the following transformations
\ba
\delta\Psi (z,\bar{z})&=&
\bar{d}\xi (z,\bar{z})  + \res{u=z}T\br{\xi (u,\bar{u}),\Psi (z,\bar{z})}
+ \ldots \nn\\
\delta\Phi (z,\bar{z})&=&\res{u=z}T\br{\xi (u,\bar{u}),{\cal T}(z)+
\Phi (z,\bar{z})}+ \ldots\nn \\
\delta\bar{\Phi}(z,\bar{z})&=&\res{u=z}T\br{\xi (u,\bar{u}),
\bar{\Phi}(z,\bar{z})}+ \ldots\label{sym}
\ea
and their conjugates
do not effect the equivalency class of deformed CFT.
Here,  we use the residue generalized for
nonholomorphic functions by the formula
$$
\res{u=z}(u-z)^k |u-z|^{2\alpha}=\Lambda(\alpha)\delta_{k+1,0} \ \ \
\br{\alpha\in\R,\ k\in\Z}\ .
 $$
where $\Lambda$ is the function used in regularization\re{reg}.

We will impose on the fields a translation invariance condition.
$$
\br{d+L_{-1}}\Xi=\br{\bar{d}+\bar{L}_{-1}}\Xi=0,\hspace{2em}
\Xi=\Psi,\Phi,\bar{\Phi}.
 $$
As regularization  used is translationally invariant, it will not reduce
the physical degrees of freedom.
The remaining symmetries correspond to translation invariant fields
$\xi$, $\bar{\xi}$.

\subsection{Linearization}
The linearized equation of motion\re{eq} and symmetries\re{sym}
can be shown respectively to be
\be{lineq}
L_1 {\cal O}_0\Psi=\bar{L}_{-1}\Phi, \hspace{2em}
\bar{L}_1\bar{\cal O}_0\Psi=L_{-1}\Phi
\ee
and
\be{linsys}
\delta\!\Psi=L_{-1}\bar{\xi}+\bar{L}_{-1}\xi, \hspace{2em}
\delta\!\Phi=L_1{\cal O}_0\xi, \hspace{2em}
\delta\!\bar{\Phi}=\bar{L}_1\bar{\cal O}_0\bar{\xi}.
\ee
Here and afterwards
$$
{\cal O}_k=\delta_{k,0}+
\sum_{j=0}^{\infty}\frac{\br{L_{-1}}^j L_{k+j}}{(k+j+1)!}\ ,
\hspace{2em}
\bar{{\cal O}_k}=\delta_{k,0}
+\sum_{j=0}^{\infty}\frac{\br{\bar{L}_{-1}}^j \bar{L}_{k+j}}{(k+j+1)!}
\ .
 $$
The equation\re{lineq} is a relaxation of the conventional closed string
equation
\be{pf}
\Phi = 0, \
\forall k\geq 0 :\
\br{L_k + \delta_{k,1}}\Psi= \br{\bar{L}_k + \delta_{k,0}}\Psi =0\ .
\ee
This relaxation is compensated by the symmetries\re{linsys}
and does not create additional physical degrees of
freedom.
The deformed representation of the Virasoro algebra in the linear
approximation can be shown as follows
\ba
L_k^{\prime}-L_k&=&\res{z=z_0}\br{z-z_0}^{k+1}\Phi(z,\bar{z})+
\res{\bar{z}=\bar{z}_0}\br{z-z_0}^{k+1}\Psi(z,\bar{z})-\nn\\
&&\res{\bar{z}=\bar{z}_0}
\sum_{j=0}^{k+1}\frac{(k+1)!}{j!}\br{z-z_0}^j{\cal O}_{k+1-j}.
\label{linrep}
\ea
In the case of the primary field solution\re{pf}, only the second term
in\re{linrep} contributes and gives a regularized version of
the deformation suggested in \cite{eo,sen}.
Calculations made up to the third order  have shown that
in the low-energy limit the theory includes
local space time symmetries and a class of solutions described by
the Brans-Dicke equation. It generalizes the analogous result \cite{eg,eg1}
 found in the
linear approximation and gives one more proof of the self-consistency of the
presented formalism.
\sect{Conclusion}
The theory presented here can be considered to be a closed string field theory,
because to its solutions  there explicitly  correspond CFT and,
in the first approximation, states of the free closed string.
Besides, the low-energy limit can be related to Brans-Dicke theory.
\renewcommand{\thesection}{}
\sect{\hspace{-1em}Acknowledgements}
Author thanks Mark Evans for inspiring him to develop the {\em deformation}
approach, for discussions and for many important suggestions.

\end{document}